\documentclass[preprint]{aastex}
\usepackage{emulateapj5,psfig}
\usepackage{apjfonts}
\tighten

\def\lta{\mathrel{\rlap{\lower 3pt\hbox{$\mathchar"218$}}
     \raise 2.0pt\hbox{$\mathchar"13C$}}}
\def\gta{\mathrel{\rlap{\lower 3pt\hbox{$\mathchar"218$}}
     \raise 2.0pt\hbox{$\mathchar"13E$}}}

\newcommand\kmsMpc{km~s$^{-1}\,$Mpc$^{-1}$}

\newcommand\hkpc{$h^{-1}\,$kpc}
\newcommand\etal{{et~al.}} 
\newcommand\mM{\ifmmode(m{-}M)\else$(m{-}M)$\fi}
\newcommand\msun{\ifmmode{\hbox{M$_\odot$}}\else{M$_\odot$}\fi}
\newcommand\sna{SNe\,Ia}
\newcommand\hst{{\it HST}}
\newcommand\kcor{$K$-correction}
\newcommand\kcors{$K$-corrections}

\slugcomment{Received 2002 December 20; accepted 2003 February 13}
\journalinfo{To appear in {\it The Astrophysical Journal}, 2003 June 1}
\shortauthors{Blakeslee et al.}
\shorttitle{ACS Supernovae}

\begin{document}

\title{Discovery of Two Distant Type Ia Supernovae in the Hubble Deep Field North
with the Advanced Camera for Surveys}

\author{
John P. Blakeslee\altaffilmark{1},
Zlatan I. Tsvetanov\altaffilmark{1},
Adam G. Riess\altaffilmark{2},
Holland C. Ford\altaffilmark{1},
Garth D. Illingworth\altaffilmark{3},
Daniel Magee\altaffilmark{3},
John L.\ Tonry\altaffilmark{4},
Narciso Ben\'\i tez\altaffilmark{1},
Mark Clampin\altaffilmark{2},
George F. Hartig\altaffilmark{2},
Gerhardt R.\ Meurer\altaffilmark{1},
Marco Sirianni\altaffilmark{1},
David R. Ardila\altaffilmark{1}
Frank Bartko\altaffilmark{5},
Rychard Bouwens\altaffilmark{3},
Tom Broadhurst\altaffilmark{6},
Nicholas Cross\altaffilmark{1},
P. D. Feldman\altaffilmark{1},
Marijn Franx\altaffilmark{7},
David A. Golimowski\altaffilmark{1},
Caryl Gronwall\altaffilmark{8},
Randy Kimble\altaffilmark{9},
John Krist\altaffilmark{2},
Andr\'e R.\ Martel\altaffilmark{1},
Felipe Menanteau\altaffilmark{1},
George Miley\altaffilmark{7},
Marc Postman\altaffilmark{2},
Piero Rosati\altaffilmark{10},
William Sparks\altaffilmark{2},
L.-G. Strolger\altaffilmark{2},
Hien D.\ Tran\altaffilmark{1},
Richard~L.~White\altaffilmark{2},
and Wei Zheng\altaffilmark{1} \vspace{0.2cm}
}

% \email{jpb@pha.jhu.edu}

\altaffiltext{1}{Department of Physics and Astronomy,
Johns Hopkins University, Baltimore, MD 21218}
\altaffiltext{2}{Space Telescope Science Institute,
3700 San Martin Drive, Baltimore, MD 21218}
\altaffiltext{3}{Lick Observatory, University of
California, Santa Cruz, CA 95064}
\altaffiltext{4}{Institute for Astronomy, University of Hawaii,
2680 Woodlawn Drive, Honolulu, HI 96822}
\altaffiltext{5}{Bartko Sci.\ \& Tech., P.O. Box 670, Mead, CO 80542-0670}
\altaffiltext{6}{The Racah Institute of Physics, Hebrew University, Jerusalem 91904, Israel}
\altaffiltext{7}{Leiden Observatory, P.O. Box 9513, 2300 Leiden, The Netherlands}
\altaffiltext{8}{Deptment of Astronomy and Astrophysics, The Pennsylvania State University,
University Park, PA 16802.}
\altaffiltext{9}{NASA-GSFC, Greenbelt, MD 20771.}
\altaffiltext{10}{European Southern Observatory, Karl-Schwarzschild-Str. 2, 
D-85748 Garching, Germany}

\begin{abstract}
We present observations of the first two supernovae discovered with the
recently installed Advanced Camera for Surveys (ACS) on the {\it Hubble
Space Telescope} (\hst).  The supernovae were found in Wide Field
Camera images of the Hubble Deep Field North taken with the F775W, F850LP,
and G800L optical elements as part of the ACS guaranteed time observation
program.  Spectra extracted from the ACS G800L grism exposures confirm
that the objects are Type~Ia supernovae (\sna) at redshifts
$z{\,=\,}0.47$ and $z{\,=\,}0.95$.  Follow-up \hst\ observations have
been conducted with ACS in F775W and F850LP and with NICMOS in the
near-infrared F110W bandpass, yielding a total of 9 flux measurements in
the 3 bandpasses over a period of 50 days in the observed frame.
We discuss many of the important issues in doing
accurate photometry with the ACS.
We analyze the multi-band light curves using two different
fitting methods to calibrate the supernovae
luminosities and place them on the \sna\ Hubble diagram.
The resulting distances are consistent with the redshift-distance 
relation of the accelerating universe model, although evolving
intergalactic grey dust remains as a less likely possibility.
The relative ease with which these \sna\ were found, confirmed, and
monitored demonstrates the potential ACS holds for revolutionizing the
field of high-redshift \sna, and therefore of testing
% and tightening the constraints on the accelerating universe cosmology.
the accelerating universe cosmology and constraining the ``epoch of deceleration.''
\end{abstract}

\section{Introduction}

The discovery of an apparent acceleration in the universal expansion
(Riess \etal\ 1998; Perlmutter \etal\ 1999) using Type~Ia supernovae
(\sna) has been one of the prime motivations for the recent
dramatic shift in the prevailing cosmological paradigm.
The formerly favored, spatially flat, matter-dominated Einstein-deSitter model
has given way to the new, observationally mandated ``concordance model'' 
possessing a higher expansion rate, a lower matter
density consistent with long-standing indications from cluster and
large-scale structure studies, and a mysterious dark energy component
usually identified with Einstein's cosmological constant~$\Lambda$.
Like the deposed paradigm, the new one is also spatially flat with a
total energy density equal to the critical density, as required by
cosmic microwave background (CMB) analyses (e.g., Jaffe \etal\ 2001;
Pryke \etal\ 2002), but the ratio of dark energy
to matter at the present epoch is roughly~2:1.

While a dark energy component is the favored explanation for the
unexpected faintness of distant \sna, other possibilities include
luminosity evolution and grey dust obscuration.  In their simplest form,
these alternative models predict that \sna\ should continue
to grow fainter with respect to a fiducial universe having
constant expansion.  On the other hand, dark energy models predict 
that \sna\ should become
differentially brighter at redshifts $z{\,\gta\,}1$ as a result of past
deceleration.  So far, only the uniquely distant SN\,1997ff at $z\sim1.7$ 
has provided evidence for past deceleration (Gilliland \etal\ 1999;
Riess \etal\ 2001), and this object was never spectroscopically confirmed as Type~Ia
(although its host and colors indicate it was).
% (although its host and colors support this inference).
Only further high-quality data on \sna\ at $z\gta1$ can confirm the cosmological 
explanation and eliminate the competing astrophysical ones. 
Such data will also help in pinning down precise values
of the cosmological parameters, which are now only weakly constrained
without the inclusion of external information, such as from the CMB.~~ 

Fortunately, the hunt for high redshift supernovae is becoming easier.
Large format CCD mosaic cameras can efficiently search over a much
wider area of the sky and spectra of faint \sna\ candidates can now 
be obtained using 8-meter class telescopes.  Another recent
development expected to benefit the field of high-$z$ \sna\ research
is the installation of the new Advanced Camera for Surveys (ACS) 
(Ford \etal\ 1998) on the 
{\it Hubble Space Telescope} (\hst).  ACS has roughly twice the field of view,
higher resolution, and nearly five times the sensitivity of WFPC2 at 8000\,\AA;
% the Wide Field Planetary Camera~2 (WFPC2); 
this combination promises to increase the discovery efficiency
of \hst\ by an order of magnitude.  

Here, we present observations of the first two \sna\ discovered with
ACS.  The objects were found serendipitously in images of the Hubble
Deep Field North (HDFN) taken by the Instrument DevelopmentTeam
(IDT) as part of the ACS guaranteed time observation (GTO) program.
The following section describes our discovery and follow-up observations,
extraction of the grism spectra, and photometric measurements.
Section~\ref{sec:distance} presents the \sna\ light curves,
the inferred distance moduli, and the updated Hubble diagram.
Section~\ref{sec:disc} then discusses the implications of our results
for cosmology
% context of published high-redshift \sna\ and the prevailing cosmological model.
%  It also briefly discusses 
and the prospects of ACS for the high-$z$ \sna\ search field.
% The final section summarizes our key results.

%%%%%%%%%%%%%%% Figures %%%%%%%%%%%%%%%%%%%5

\begin{figure*}\epsscale{1.05}
%\plotone{sn2002dc_mosaiclet.eps}
\plotone{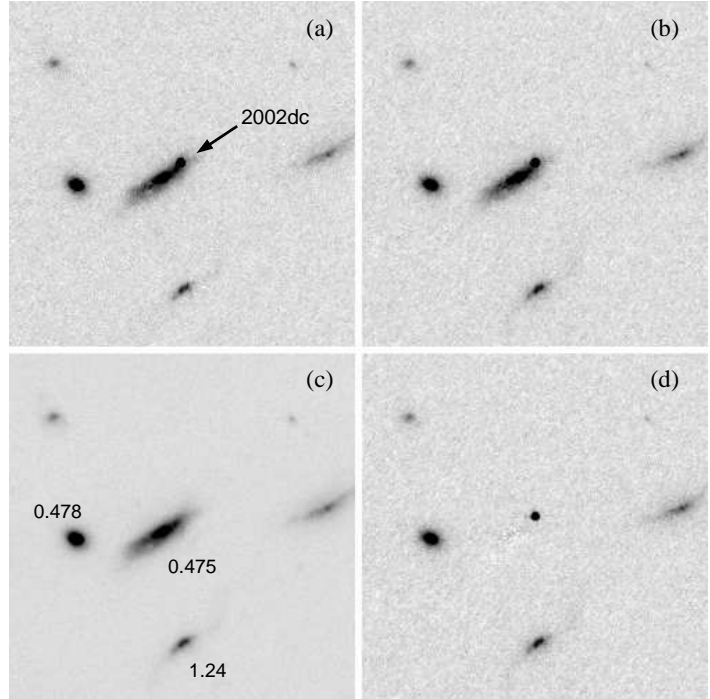}\medskip
\caption{Comparison of a small region of our ACS images centered on the position
of SN2002dc with the same region from the original 1995 WFPC2/F814W image of the field.
The displayed field size is $12\farcs5{\times}12\farcs5$ (250 ACS WFC pixels on a side)
and the position angle is $140^\circ$.  The separate panels show:
ACS/F775W image (a); ACS/F850LP image (b);
original WFPC2/F814W image (c); ACS F850LP image with the 
scaled WFPC2/F814W image of the SN2002dc host galaxy subtracted (d).
The ACS images are displayed without any relative scaling of the counts,
showing the closely comparable signal levels in the two bands.
The three galaxies in this field with measured redshifts (Cohen \etal\ 2000)
are labeled in the lower left (WFPC2) panel, including the $z{\,=\,}0.475$ 
host galaxy. SN2002dc is immediately apparent after subtraction of the host.
\label{fig:dcstamps}}
\end{figure*}

\begin{figure*}
%\plotone{sn2002dd_mosaic4let.eps}
\plotone{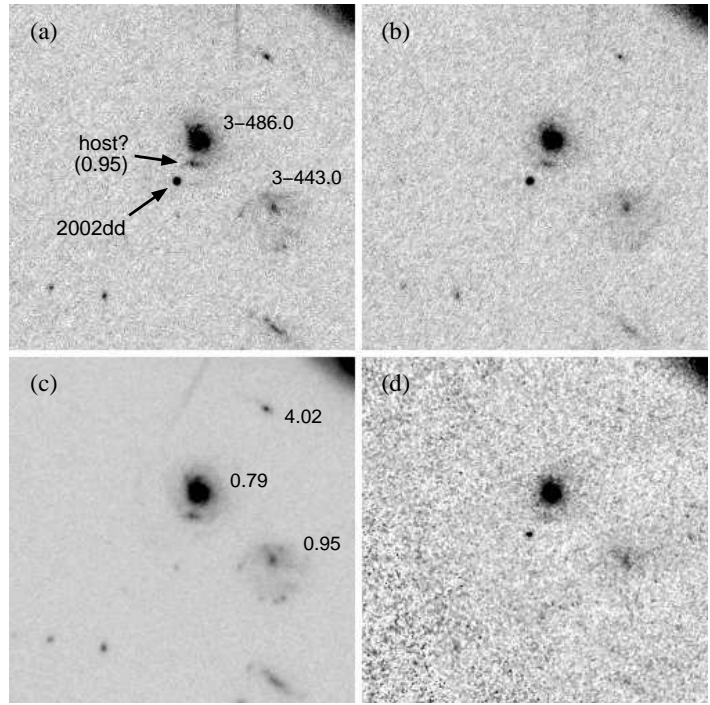}\medskip
\caption{Comparison of our ACS and NICMOS images centered near SN2002dd 
with the same region from the original 1995 WFPC2/F814W image of the field.
The displayed field size and orientation is the same as Fig.\,\ref{fig:dcstamps}.
The panels show: ACS/F775 image (a); ACS/F850LP image (b);
original (1995) WFPC2/F814W image (c); follow-up NICMOS/F110W image
one month after discovery.
Galaxies in this field with measured redshifts (Cohen \etal\ 2000)
are labeled in the lower left (WFPC2) panel.
The candidate host galaxy identified in the text is labeled in the
upper left panel; it has AB magnitudes $\hbox{F775W}{\,\approx\,}25.5$,
$\hbox{F850LP}{\,\approx\,}25.3$.
\label{fig:ddstamps}}
\end{figure*}

\section{Observations and Data Reductions}

%\subsection{Direct Imaging}
%\subsection{Image Processing}
\subsection{Imaging Data}

The Hubble Deep Field North (HDFN) was observed with the ACS Wide Field
Camera (WFC) as part of the GTO program, proposal number 9301.  
The observations were conducted on 2002 May 11 in the F775W (SDSS $i$) and
G800L (grism) bandpasses for two and three orbits, respectively, 
and on 2002 May 21 in F850LP (SDSS $z$) for three orbits.
A complete description of the GTO program will be given in a future paper;
here we concentrate on our measurements of the two supernovae 
that were found in these data.

The images were initially processed through the STScI CALACS 
pipeline (Hack 1999),  which performs the bias and dark subtraction,
flat-fielding, and conversion of the counts to electrons.
Further reduction was done through the ``Apsis'' GTO data pipeline,
described  in detail by Blakeslee \etal\ (2002).  Apsis measures offsets,
rejects cosmic rays (CRs) and detector defects, and combines the images to a 
single geometrically corrected one using the drizzle method
(Fruchter \& Hook 2002).  We used the default ``square'' kernel 
(bilinear interpolation) with an output scale of 0\farcs05/pix for
the drizzling.  The resulting full width at half maximum (FWHM) of the point
spread function (PSF) was about 0\farcs104 in F775W and 0\farcs110 in F850LP.
While alternate kernels or output scales could reduce the PSF FWHM,
this approach allows us to apply directly the aperture corrections
from the ACS photometric calibration program (Sirianni \etal\ 2003).

Inspection of the reduced images revealed two bright point sources
that were not present in the original 1995 HDFN images.
Figures~\ref{fig:dcstamps} and \ref{fig:ddstamps} show subsections of the 
F775W and F850LP ACS images centered on these objects,
as compared to the same regions taken from the 1995 WFPC2/F814W images.
The first object, designated SN2002dc (Magee \etal\ 2002), is located at
$\hbox{R.A.} = 12^{\rm h}36^{\rm m}49\fs84$,
$\hbox{Dec.} = {+}62^\circ13^\prime13\farcs0$ (J2000), where
our coordinates are on the system of the Fern{\' a}ndez-Soto \etal\ (1999)
catalogue.  SN2002dc is 0\farcs84 from the center of its apparent host 
galaxy, which is number 2-264.1 in the Williams \etal\ (1996) HDFN Catalog
and has a measured redshift $z=0.475$ (Cohen \etal\ 1996).

The second object, SN2002dd (Tsvetanov \etal\ 2002), was originally
identified by its spectrum; it has coordinates
$\hbox{R.A.} = 12^{\rm h}36^{\rm m} 55\fs36$,
$\hbox{Dec.} = {+}62^\circ12^\prime46\farcs1$ (J2000).
It is located 1\farcs64 from the center of the HDFN Catalog galaxy 3-486.0;
however, this is not the host since it has a redshift $z=0.79$ 
(Cohen \etal\ 1996), whereas the SN2002dd spectrum (see below) places
it at $z\approx0.95$ (note that the 11\% higher redshift from the discovery
telegram was based on the rough ground-based calibration of the ACS grism). 
Interestingly, the diffuse galaxy 3-443.0 has $z=0.950$ (Cohen \etal\ 2000)
and is centered 3\farcs6 away (projected separation $\lta 20$ \hkpc).
In addition, there is a fainter, uncatalogued object 0\farcs87 (4.7 \hkpc)
from SN2002dd projected {\it between} it and 3-486.0 (see Figure~\ref{fig:ddstamps}).
This object's grism spectrum is
consistent with a starburst galaxy at $z{\,=\,}0.95$
(see the following section), and there is some indication of faint
luminosity (possibly a spiral arm) arching from it towards
SN2002dd. We consider this the most likely host galaxy.

Because the ACS grism spectra indicated that these
objects were both \sna\ near maximum luminosity, we reobserved this field
on 13~June under HST GO proposal \#9352 for one ACS orbit split between 
F775W and F850LP and one orbit with Camera~2 of the recently revived
NICMOS instrument in the F110W bandpass (SN2002dd only).  
Two additional single-orbit ACS observations done
under GTO proposal \#9301 on 24~June (split between F775W and F850LP),
and 30~June (F850LP).   These follow-up images were
processed in the identical manner as the original HDFN GTO data,
except that they were CR-split and CALACS performed the CR rejection.
The NICMOS images were processed through
the STScI CALNIC software and then combined using drizzle.
% Near-Infrared Camera and Multi-Object Spectrometer Observations 

\subsection{Grism Spectra}
%\section{Grism Spectra}

Preliminary discussions of the ACS G800L grism performance are given by Walsh
\etal\ (2002) and Ford \etal\ (2002).  The spectral dispersion for WFC
grism exposures is 40\,\AA\,pix$^{-1}$ on average, but varies with position by
$\pm10$\%. The resolution is typically $\sim\,$90\,\AA.
Spectra of the two HDFN supernovae were extracted from the reduced 
ACS/WFC grism images using the aXe software (Pirzkal \etal\ 2002),
which also performs the wavelength and flux calibrations.  
The on-orbit wavelength calibration is accurate to $\lta\,$10\,\AA,
or $\sim\,$0.001 in redshift, and has been
discussed in detail by Pasquali \etal\ (2002).
We subtracted a polynomial fit to the sky in the cross-dispersion
direction in the 2-d image before aXe extraction.  For SN2002dc,
we also needed to fit and remove the contribution from the host
galaxy, so we masked the region of extraction and fit a polynomial
to the very close background, interpolating over the masked region.
The resulting local background was very flat.

We cross-correlated the extracted spectra against a database of
high-signal-to-noise spectra of 27 different \sna, most of which had
data from multiple different dates.  We determine the redshift 
from each template that gave a significant and unambiguous
correlation peak, taking the rms dispersion in redshift values
as an estimate of the uncertainty. 
We also performed cross-correlations against spectral templates for
other supernovae types (Filippenko 1997), but none of these yielded
an acceptable fit to either of the spectra.
For SN2002dc, we find excellent matches with normal \sna\ such
as SN1995D, SN1994ae, SN1995E, and SN1994S.  These overlap more
than half of the rest-frame spectral range and give
cross-correlation $r$~values (Tonry \& Davis 1979) of 16 or more.
The best consistency is for an age $t=1\pm3$ days past maximum,
and the best-fit redshift is $z{\,=\,}0.473\pm0.006$, with a total
range from 0.45 to 0.49.  This is in excellent agreement with the
$z=0.475$ redshift for the host galaxy from Cohen \etal\ (1996, 2000).

\psfig{file=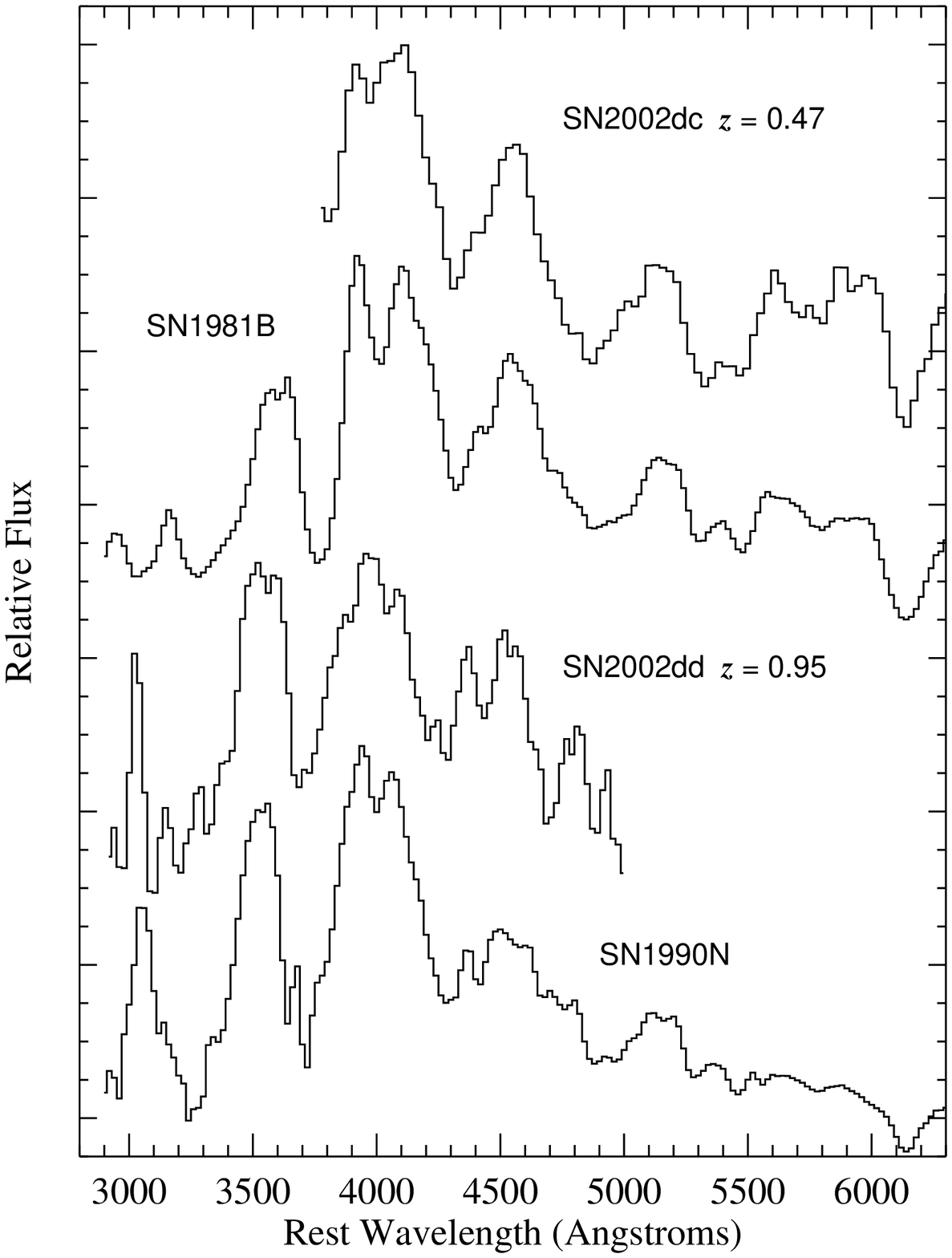,width=9.1cm}
\figcaption{ACS grism spectra transformed to the rest frame for SN2002dc at $z{=}0.47$ 
and SN2002dd at $z{=}0.95$ are compared with ground-based spectra from two
redshift zero \sna\ near maximum luminosity.  The SN2002dc spectrum closely 
resembles that of SN1981B in the galaxy NGC\,4536, including the 
Si\,{\sc ii} absorption at 6150\,\AA, which is the defining feature of \sna.
This is the first time this feature has been seen at $z\approx0.5$.
The spectrum of SN2002dd appears intermediate between those of SN1981B and
SN1990N in NGC\,4639.  In particular, it shows the deep Ca\,{\sc ii} trough 
at 3750\,\AA\  and numerous other features due to 
Si\,{\sc ii} and Fe\,{\sc ii} common to \sna.
\label{fig:snspect}}
\par\vspace{1.25cm}

\psfig{file=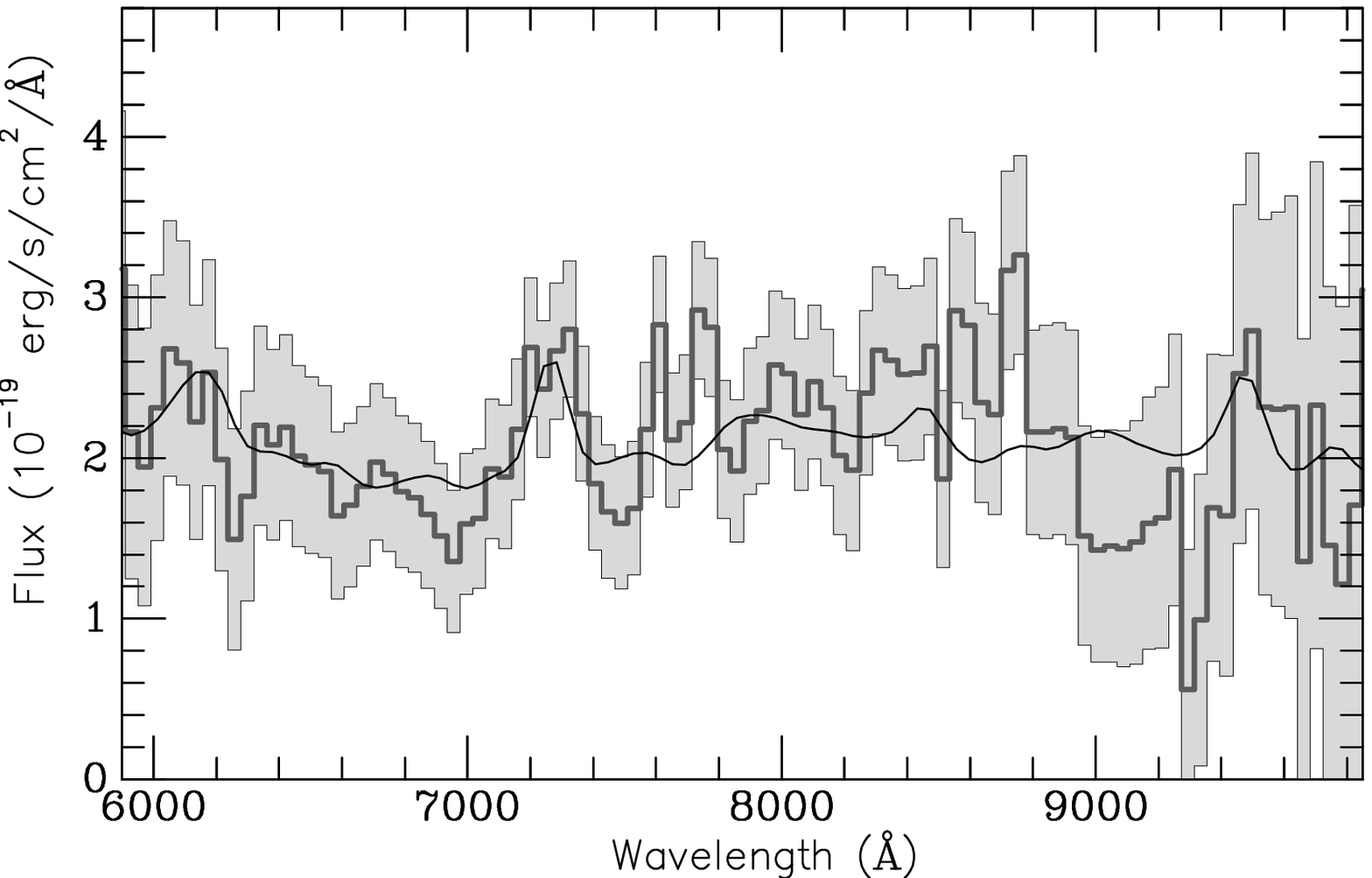,width=8.75cm}\vspace{0.2cm}
\figcaption{Observed ACS grism spectrum of the SN2002dd candidate host 
galaxy (see Fig.\,\ref{fig:ddstamps}). The 1-$\sigma$ error region is
delineated by the gray area.  The dark curve 
% shown for comparison 
is a spectrum of a nearby starburst galaxy from Calzetti \etal\ (1994)
%% http://www.stsci.edu/ftp/catalogs/nearby_gal/sed.html
that has been redshifted to $z{\,=\,}0.95$, smoothed to the
resolution of the ACS grism extraction, and rescaled for comparison
purposes.  The broad line near
7200\,\AA\ in the grism spectrum aligns well with the 
[O\,{\sc ii}] $\lambda3727$ emission in the comparison spectrum
for this redshift.
\label{fig:hostspect}}
\par\vspace{0.6cm}

For SN2002dd, we find good spectral matches with SN1995D, SN1990O, SN1981B,
SN1972E, SN1990N, and SN1994ae, among others.  The range of acceptable ages 
is fairly broad, $-$7 to $+$9 days past maximum.  
The best-fit redshift is $z{\,=\,}0.944\pm0.022$, with a total range of 0.90--0.97.
These results rule out the $z=0.79$ galaxy 1\farcs6 away as a possible host,
but are consistent with the $z=0.950$ redshift of the galaxy 3\farcs6 away 
(our grism spectra for both these galaxies support the results from Cohen \etal).
We note that a number of spectral templates,
including SN1989B, SN1992a, SN1994m, and SN1996x, produced
formally acceptable fits, but with $z{\,\approx\,}0.5$.
In fact, many of the templates have some correlation peak at this
redshift, but these fits are all of significantly lower quality 
than the best fits with $z{\,\approx\,}0.94$.  
In addition, the photometric data for SN2002dd is completely
incompatible with $z{\,\approx\,}0.5$.
While the correct redshift is clear in this case, we wish to caution
that \sna\ cross-correlation results in general depend critically 
on the wavelength coverage and the set of spectral templates used.

Figure~\ref{fig:snspect} shows the calibrated, rest-frame spectra of 
the supernovae with a 5-pixel (2--2.5 times the FWHM) extraction
aperture, and compares them to smoothed
spectra of two nearby ``standard'' \sna\ at maximum luminosity.  
Both SN2002dc and SN2002dd display many of the standard spectral
characteristics of \sna, which have been discussed in detail by Coil
\etal\ (2000) for \sna\ at these redshifts.
In particular, we note that SN2002dc
shows the Si\,{\sc ii} absorption trough at rest $\lambda = 6150$\,\AA, which
is considered the defining feature of \sna\ and thus confirms the Type Ia 
identification.  This is the first time the complete feature has been 
unambiguously seen at $z\approx0.5$. It is nearly impossible to observe 
from the ground at these redshifts, being an absorption 
feature in a faint spectrum at $\lambda\gta 9000$\,\AA.

While the SN2002dd spectrum does not extend to the Si\,{\sc ii}
$\lambda 6150$ feature, its general characteristics easily rule out 
all other types of supernovae except the rare luminous Type Ic (see
Clocchiatti \etal\ 2000).  However, as discussed by Riess \etal\ (1998),
Type Ic supernovae tend to lack the very deep Ca\,{\sc ii} $\lambda
3750$ absorption with strong recovery to the peak at $\sim\,$4000\,\AA.
In addition, only \sna\ spectra show the notch 
% due to Si\,{\sc ii} in the peak at 4000\,\AA.  
in this peak due to Si\,{\sc ii} $\lambda4130$.
Clearly, SN2002dd shares these characteristic of \sna.
We therefore regard both objects as spectroscopically confirmed \sna\
having redshifts consistent with the host or neighboring galaxies.

Finally,  Figure~\ref{fig:hostspect} presents the
grism spectrum of the faint $i_{\rm AB}{\,\approx\,}25.5$
candidate host galaxy identified in Figure~\ref{fig:ddstamps},
and compares it to that of the starburst galaxy NGC\,3049 from Calzetti \etal\
(1994)\footnote{http://www.stsci.edu/ftp/catalogs/nearby\_gal/sed.html}.
The starburst spectrum has been redshifted to $z{\,=\,}0.95$ and smoothed to the
resolution of the grism spectrum.  The feature at $\sim\,$7200 \AA\
in the candidate host's spectrum matches the [O\,{\sc ii}] $\lambda3727$
line at this redshift. The continuum shape is also a reasonable
match, although this varies considerably among starburst galaxies.
We suggest that SN2002dd lies in the outskirts of this faint galaxy,
which in turn is a dwarf companion of the larger $z=0.95$ galaxy.

\subsection{Photometry}

%%% (1) Aperture photometry
We measured the fluxes of the two supernovae in each bandpass at
each epoch using the Vista ``psf'' routine, which derives the
scale factors between faint point sources and brighter PSF
template stars using an implementation of the DoPhot 
(Shechter \etal\ 1993) ``wingy Gaussian'' fitting function.  
Seven different high signal-to-noise (S/N) stars in the field were
used as psf templates for the F850LP images, and five of these
were also used as templates for the F775W photometry (the two
brightest were just saturated in the first epoch F775W image).
The aperture flux of the faint source is derived from the aperture flux of
the high-S/N template times the fitted scale factor.  The total flux was
then derived using the bandpass-dependent aperture corrections of Sirianni
\etal\ (2003), plus small adjustments for the color-dependence of
the F850LP aperture correction (Gilliland \& Riess 2002).
We experimented with several different
aperture sizes for the PSF stars but always derived consistent total
magnitudes at the 1\% level after applying these aperture corrections.  In
the end, we chose an aperture of radius 0\farcs5 as affording the best
compromise between systematic and random error.
Although the F850LP correction becomes very sensitive to color
at small radii for extremely red objects, the adjustment 
from Gilliland \& Riess  for objects with 
the colors of the SNe and PSF stars is only $\sim\,$0.01 mag for this aperture.
% The correction for this aperture was 0.077 mag in F775W and 0.116 in F850LP.
The adopted aperture corrections are then 0.077 mag in F775W and 0.116 in F850LP.

For SN2002dc, it was necessary to subtract the host galaxy.  
We used the very deep WFPC2/F814W image remapped to the scale and
orientation of the ACS frames and rescaled the galaxy flux to
minimize the residuals upon subtraction of the galaxy from each
of the F775W and F850LP discovery images.  The flux ratios were
determined after smoothing the ACS images to the WFPC2 resolution,
but the photometric measurements were done after subtracting the
rescaled galaxy from the original (unsmoothed) ACS images. 
Figure~\ref{fig:dcstamps} illustrates our F850LP subtraction.
Scaling of the flux to the later (shorter) observations was done according
to the exposure time.  We varied the derived scale factors by
$\pm0.5$\% (about twice the uncertainty of the flux ratio 
determinations) and repeated the SN2002dc flux measurements.
The results differed by less than 0.05\%, primarily because
the psf routine always subtracts a locally determined sky value.
Therefore, the host galaxy subtraction is not a significant source 
of error.

We empirically determined the magnitude errors and biases via
Monte Carlo experiments.  We constructed a composite
PSF star for each bandpass and added 50-100 cloned versions of this star
scaled to the measured magnitudes of each of the two supernovae.  We then
recovered these artificially introduced stars, measured their magnitudes
with the psf routine described above, and compared to the true magnitude.
We repeated this procedure 10 times at each epoch for both F775W and F850LP.
The errors determined in this way agreed closely with the errors we
estimated from first principles based on all sources of noise, including
the scatter in the results arising from different PSF templates.
The estimated and measured errors differed on average by about 0.01 mag,
and never by more than 0.02 mag.  We found some small biases in the
fitted scale factors, ranging between zero and 0.06 mag (typically half the 
measured 1-$\sigma$ error or less), and corrections for these were applied.
We also tested our approach against results obtained using DAOPHOT
(Stetson 1987), and again found agreement at the 0.01 mag level.

For the single-epoch NICMOS/F110W observation of SN2002dd, we
took a slightly different approach.  The small field of view 
%%% meant that there were 
encompassed essentially no good PSF comparison stars.  So,
instead, we created ten PSFs for this camera using Tiny\,Tim 
(Krist \& Hook 2001)
and added them to each calibrated image after determining the offsets
but prior to drizzle combination. 
We then determined the relative scale between SN2002dd and these
artificial PSFs in the drizzled image using both DAOPHOT and the
Vista/DoPhot PSF-fitting approach.

The ACS instrumental magnitudes were then transformed to the AB system
using the latest calibrations.  This is the same as derived from the
header photometric keywords for F775W, but we anticipate a correction 
of 0.064 mag to the current (2002 December) F850LP header calibration
(Sirianni \etal\ 2003).  The adopted zero points for 1$\,e^-\,$sec$^{-1}$
are then 25.654 (F775W) and 24.850 (F850LP)\footnote{The 2003 February
photometric recalibration of ACS differs from these zero points by 
$-$0.013\,mag in F775W and $-$0.006\,mag in F850LP.  These 
differences are too small to affect any of our conclusions.}.
For the light curve analysis, we require Vega magnitudes, and
use AB to Vega conversions of $-0.401$ (F775W) and $-0.569$ (F850LP).
For the NICMOS/F110W data, we calibrated directly to the Vega system using
the preliminary post-servicing mission zero point for camera~2, which is 
about $55\pm5$\% more sensitive than for cycle~7
(M.\,Dickinson \& M.\,Rieke, priv.\ comm.).  
Because of the still-tentative nature of this calibration, we have
added a conservative 0.07\,mag in quadrature to the
F110W measurement error to get the final uncertainty.

Finally, because the photometric record is most complete in F850LP but the first
F850LP observation (``epoch~2'') occurred 10 days after the first F775W and
G800L observations (``epoch~1''), we derived a first epoch F850LP magnitude from
the grism spectrum and F775W magnitude.  To do this, we used the extracted count
spectrum, divided out the G800L response function, integrated the result across
the filter bandpasses, and applied the AB zero points to derive a color directly
from the grism spectrum, $(i{-}z)_{\rm gr}$.  Because of aperture and other
effects, this will not exactly correspond to the colors measured from the direct
images. However, we plotted the epoch~2 F850LP magnitudes against the epoch~1
F775W${\,-\,}(i{-}z)_{\rm gr}$ measurements for all isolated, unsaturated point
sources having AB mag $<$ 25.  The result is shown in Figure~\ref{fig:zcal}.

A total of 21 point sources were
used to define the linear relation in Figure~\ref{fig:zcal}, 
which has a slope of 0.98 and a scatter of 0.045 mag.  
The residuals about the relation show no correlation with
$(i{-}z)_{\rm gr}$ color.  Forcing a slope of unity increases the scatter by
0.009 mag.  Both supernovae lie $\sim\,$0.2 mag above the relation, indicating
that they each faded by about this amount in the 10 day interval.  One other
relatively faint object (shown with an open circle in Figure~\ref{fig:zcal})
appears to have faded by a similar amount and was excluded from the fit, being a
$>4$-$\sigma$ outlier, although including it does not significantly change the
results.  We use the fitted relation to ``predict'' the epoch~1 F850LP
magnitudes of the \sna, and include a 0.05 mag error in quadrature to allow for
the scatter about the fit.  Our final set of photometric measurements
transformed to the Vega system are presented in Table~\ref{tab:mags}.~

\vbox{\vspace{0.4cm}\psfig{file=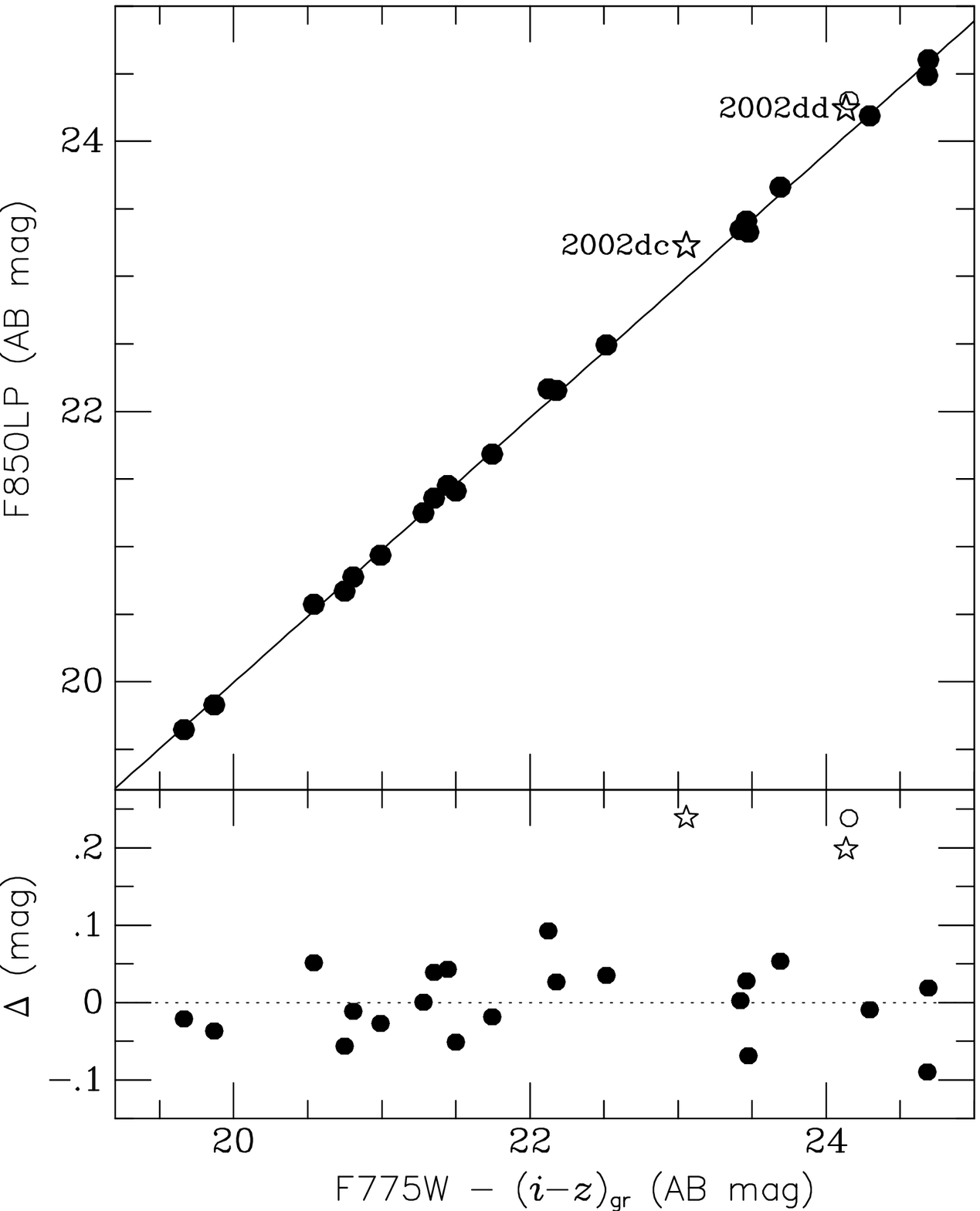,width=8.5cm}\vspace{0.2cm}}
\figcaption{Derivation of the first epoch F850LP \sna\ magnitudes.
{\it Top:} Second-epoch F850LP aperture magnitudes are
plotted against the difference of the first-epoch F775W aperture
magnitudes and $(i{-}z)_{\rm gr}$ color measured by integration 
of the grism spectra.  Magnitudes are on the AB system. 
The solid line was fitted to the 21 solid points and is used to estimate what
the F850LP magnitudes of the two supernovae would have been at the first epoch;
the offset of the supernovae above the relation indicates that
they both faded by about 0.2 mag in the 10 day interval between epochs.
One other object, shown by the open circle, appears to have gotten
fainter by a similar amount and was not used in the fit.
{\it Bottom:} Magnitude residuals about the fit.
% ; the supernovae and the outlier are off scale in this plot.
The scatter of the solid points is 0.045 mag.
\label{fig:zcal}}
\medskip

%%%%%%%%%%%%%%%%%%%%%%%%%%%%%%%%%%%%%%%%%%%%%%%%%%%%%%%%%%%%%%%%%%%%%%%

\section{Light Curves and Distances}
\label{sec:distance}

From the multicolor light curves defined by Table~\ref{tab:mags}, we
can calibrate the \sna\ luminosities, and so infer their distances.
Several different methods exist for doing this,  including the 
$\Delta m_{15}$ decline rate parameterization (Phillips 1993),
$\Delta m_{15}$ template fitting (Hamuy \etal\ 1995, 1996),
multi-color light curve shape (MLCS) (Riess \etal\ 1996, 1998), 
stretch (Perlmutter \etal\ 1997), 
and best average template (BAT) (Tonry \etal\ 2003) methods.
Here we employ both the MLCS and BAT methods to estimate the distances
to SN2002dc and SN2002dd and place them on the Hubble diagram.
BAT, while less developed than MLCS, does not assume a
single-parameter family of objects, and therefore we can use it
to test for consistency between two significantly different approaches 
to measuring \sna\ distances.

\subsection{$K$-corrections}

All methods for estimating \sna\ distances depend on how well the data on
the distant program objects and the calibrating set of local \sna\ can be
transformed to a common reference frame and photometric system for comparison.
This can be achieved either by transforming the magnitudes of the distant 
objects to the closest-matching rest-frame bandpass in which
the calibrating sample have been measured, or equivalently, through redshifting
the photometric record of the calibrators and transforming to the bandpasses
in which the distant \sna\ were measured.  In either case, 
the systematic error is minimized if the observed and template bandpasses
match as closely as possible after shifting to the common reference frame.
We note that the MLCS and BAT methods used in the analysis below take
the opposite approaches as far as redshifting/blueshifting the
calibrator/program objects.

Because the observation and comparison bandpasses will not match perfectly even
after redshifting to a common frame, it is necessary to apply ``\kcors'' to
convert the observed magnitudes to the template bandpasses.  The problem of
\kcors\ has been discussed in the context of \sna\ by Hamuy \etal\ (1993); Kim
\etal\ (1996); Schmidt \etal\ (1998); and Nugent \etal\ (2002), among others.
We calculate the \kcors\ in a manner similar to Riess \etal\ (1998) and
Tonry \etal\ (2003), except we
use the model spectral energy distributions (SEDs) from Nugent \etal\ (2002).
However, we force the models to agree with the average \sna\ light curve
from Jha (2002), which is based on extensive empirical multi-color data.

For each day that the model and template light
curve data are tabulated, we integrate the model SEDs through the $U,B,V,R,I$
bandpasses, fit a spline to the differences with respect to the empirical light
curve, multiply the result into the SED, and iterate to convergence.  Once the
model SED colors have converged to the observed light curve colors for every date, 
they can be used for the \kcor\ calculation.    In this way, we preserve
both the detailed spectral features of the models and the general
continuum shape defined by the empirical light curves.
We include an allowance for $\pm0.07$ mag error in the \kcors\ based on the 
scatter derived from individual
\sna\ near maximum-light (B.\,Schmidt, priv.\ comm.).  This method
ensures consistency with previous \sna\ distance analyses while allowing
us to obtain reliable $K$-corrections for our rest-frame $U$-band data.
We note that if there is significant uncertainty in the redshift, this 
should also be included in the error analysis, but the \kcor\ near maximum
is not highly sensitive to the precise redshift.  Also, in the present 
case, we believe that the \sna\ redshifts are accurate to 0.01 (better
for SN2002dc), based on a combination of the \sna\ and galaxy spectra.
Table~\ref{tab:mags} gives the F850LP \kcors\ used in the MLCS analysis.

%%%%%%%%%%%%%%%%%%%%%%%%%%%%%%%%%%%%%%%%%%%%%%%%%%%%%%%%%%%%

\subsection{Distances} % & extinctions

We estimate the \sna\ distances using both the MLCS method described in
detail by Riess \etal\ (1996) and developed further by Riess \etal\ (1998),
and the BAT method described by Tonry \etal\ (2003).   Both methods use
libraries of multiband \sna\ light curves.  The main difference is
that MLCS parameterizes the light curve shape as a continuous 1-parameter function
of the luminosity excess $\Delta$, whereas BAT determines the \sna\ luminosity
from a weighted average among the best-matching comparison light curves
and uses the scatter in its error estimate.
Both methods fit for the time of maximum and yield estimates of the
extinction based on the color difference between model and observations.
The distance is then inferred from the calibrated light curve after 
applying the derived rest-frame extinction correction.
This should not be confused with the extinction correction of
$A_V = 0.04$ mag (Schlegel \etal\ 1998) that we apply first to
the observed magnitudes prior to either method.

Table~\ref{tab:results} lists the results of our MLCS and BAT analyses.
The columns are: (1) \sna\ designation; (2) redshift;
(3) MLCS luminosity parameter $\Delta$ (mag); (4) MLCS estimated
distance modulus; (5) MLCS fitted $V$-band extinction (mag); 
(6) best-matching \sna\ light curve template from the BAT method;
(7) BAT estimated distance modulus; (8) BAT fitted $V$-band extinction (mag).
The sense of the $\Delta$ parameter from the MLCS fit is that SN2002dc
is found to be underluminous by 0.1 mag, while SN2002dd is overluminous
by 0.4 mag (i.e., systematic errors of these sizes would be made in
assuming that \sna\ were perfect standard candles).
The light curve fits for SN2002dd were done both with and without
the F775 data, corresponding to the rest-frame $U$ band.
The two methods produce very similar results in each case,
well within the quoted 1-$\sigma$ fit errors.
Although SN2002dc is nearer and brighter, the estimated
errors from both methods are larger for this object because it
is found to have significantly higher extinction, which adds to the
distance uncertainty.  The extinction is consistent with the
presence of SN2002dc within the disk of its host galaxy.

\psfig{file=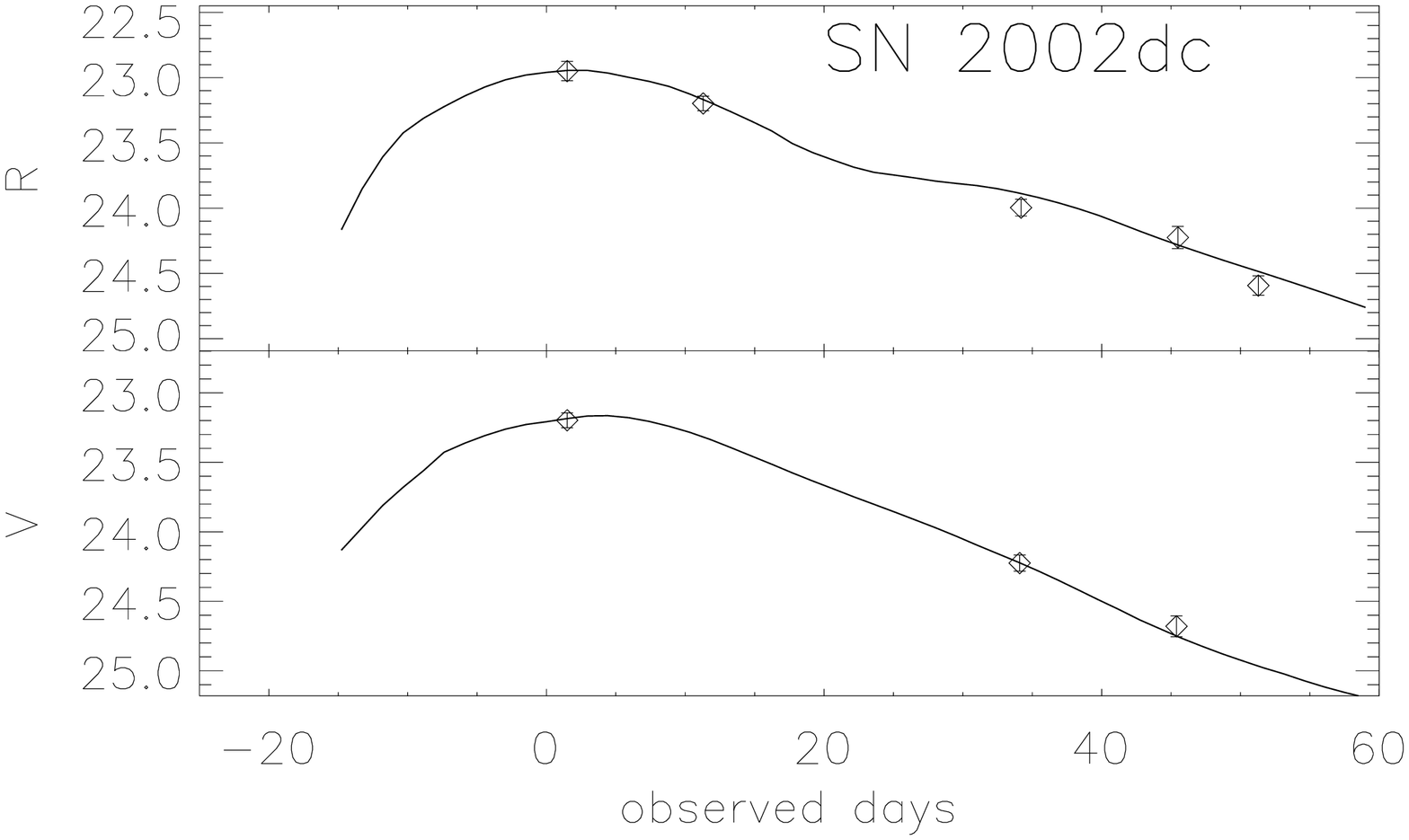,width=8.8cm}
\figcaption{Light curves and fitted MLCS model for SN2002dc in 
the rest-frame $V$ and $R$ bands.  The model
includes: luminosity parameter $\Delta$, which determines the
shape of the curves; reddening, which sets the extinction;
time of maximum; and the distance.  All points were
used in deriving the fit.
\label{fig:dc_lc}}
\vspace{0.5cm}

\psfig{file=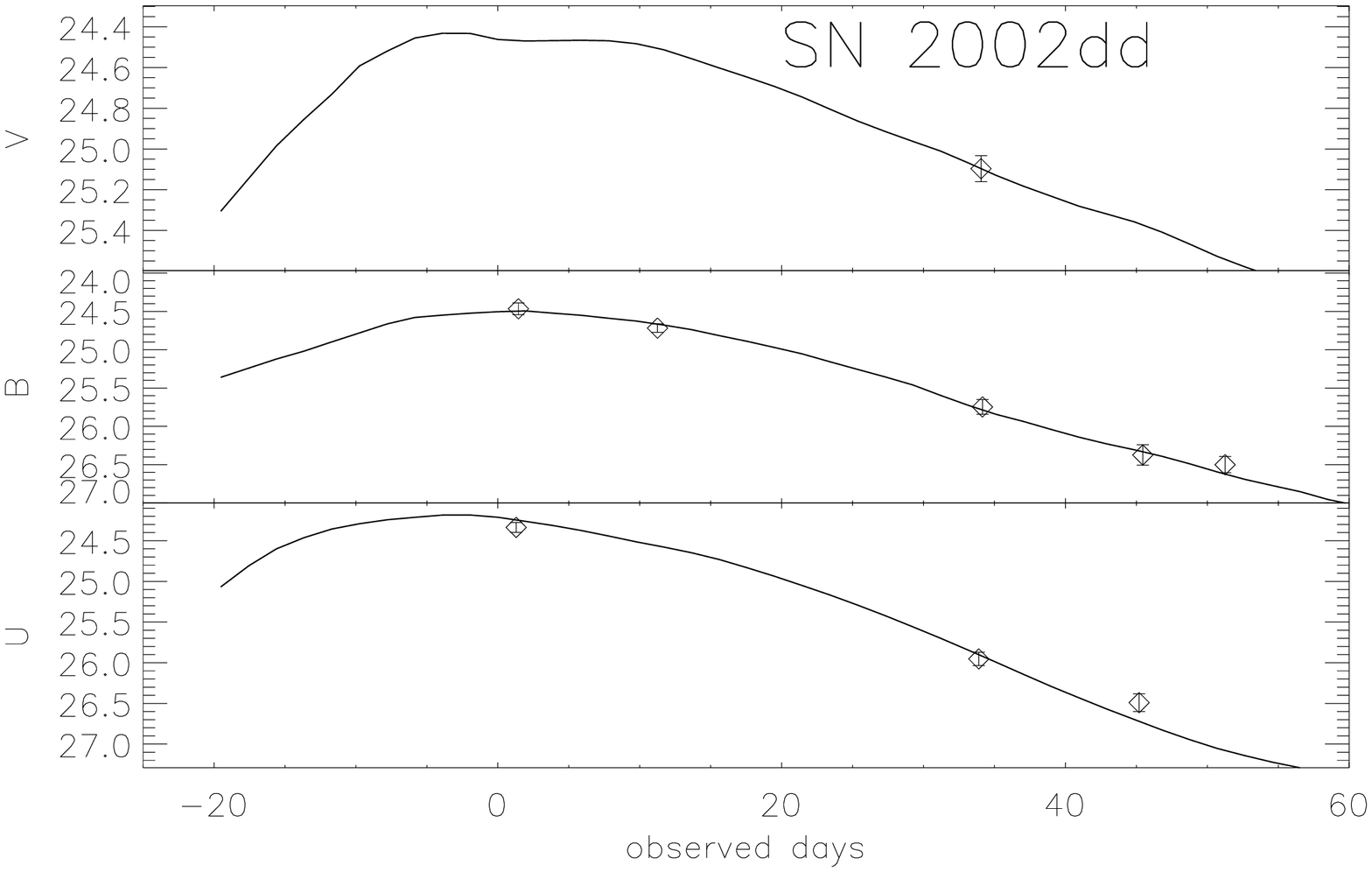,width=8.8cm}
\figcaption{Light curves and fitted MLCS model for SN2002dd.  Only 
the F850LP and F110W measurements (rest-frame $B$ and $V$, respectively)
were used in deriving this fit. The rest-frame $U$ data (observed F775W)
points were not used, but are shown in comparison to the $U$ light
curve of the fitted model.  The model appears to decline more rapidly
than the data in $U$, which could be due to metallicity or
other effects that become much more pronounced in the ultraviolet,
or due to insufficient $U$-band template data. \\
\label{fig:dd_lc}}\vspace{0.3cm}

The light curve fits for SN2002dc and SN2002dd from the MLCS method
are shown in Figures~\ref{fig:dc_lc} and \ref{fig:dd_lc}, respectively.
Overall, the fits look quite good.  
For SN2002dc, the observed F775W and F850LP observations are transformed
to rest-frame $V$ and $R$, which provide the best matches at this redshift.
The results are basically unchanged if the transformation is instead 
done to $B$ and $V$, which is an indication of the consistency of
the $K$-corrections.  The $R$-band light curve (observed F850LP) shows 
a hint of the late-time ingress of flux at red wavelengths, a
common characteristic of \sna.

For SN2002dd, the F850LP and F110W bandpasses transform most readily 
to rest-frame $B$ and $V$, which
are the most commonly used bandpasses for fitting \sna\ light curves.
% Comparatively little data is available on $U$ light curves. 
The empirical library of $U$-band light curve data is much less complete.
Moreover, the shape of the UV continuum becomes sensitive to the metallicity
(e.g., H\"oflich \etal\ 1998), 
which can cause systematic errors in the fits.
%  Moreover, The empirical library of UV light curve data is much less complete.
These problems are compounded by the dust obscuration being
greater in $U$, so the peculiar systematic effects of this 
poorly constrained bandpass can unduly drive the fitted extinction values.
For these reasons, rest-frame UV data have not in the past been used 
for light curve fitting, and our results omitting the F775W data may
be more reliable; however, the analysis is not overly sensitive to
their inclusion.  Figure~\ref{fig:dd_lc} shows the MLCS model from
the fit to the F850LP and F110W data alone, but compares to
the full data set.  The agreement is reasonably good in $U$, 
indicating that SN2002dd is not particularly abnormal in the UV,
although the data appear to 
decline more slowly than the model, possibly due to the 
effects mentioned above.
A better understanding and more extensive UV light curve data sets
are needed to take full advantage of optical observations of \sna\
at $z{\,\gta\,}1$.

\subsection{The Hubble Diagram}

\begin{figure*}\epsscale{1.3}
%\plotone{sncosmo.eps}
\vspace{0.3cm}
\plotone{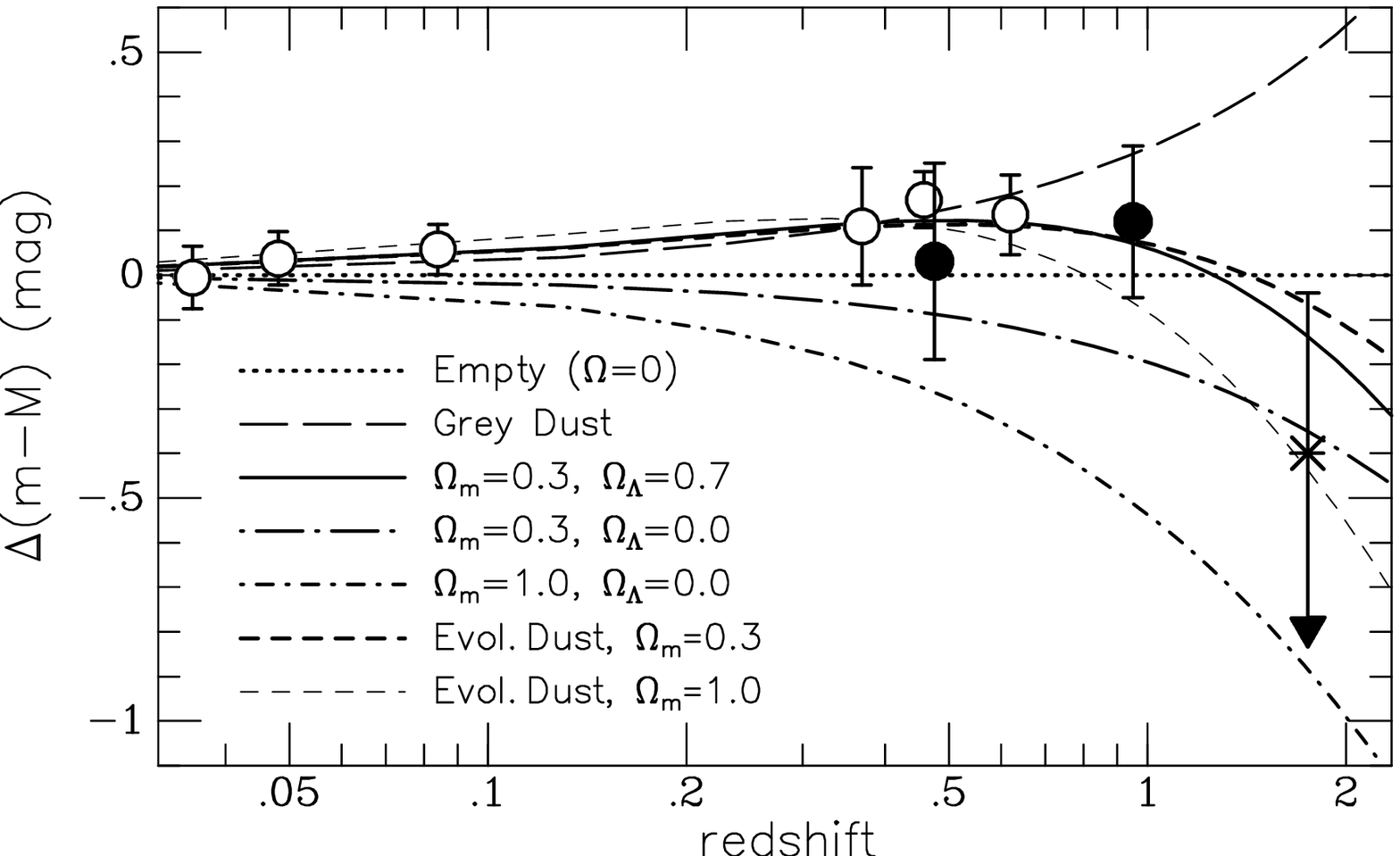}\vspace{0.1cm}
\caption{The differential Hubble diagram of previously published and new 
\sna\ compared to various cosmological models, with and without
grey dust obscuration. 
SN2002dc and SN2002dd from this work are shown with solid symbols
(MLCS distances and errorbars).
The previously published, mainly ground-based data 
have been grouped into six redshift bins as in Riess \etal\ (2001)
and are shown with open circles.
The original $z{\,<\,}0.1$ data come from Hamuy \etal\ (1996) 
and Riess \etal\ (1999), while the $0.3 \lta z \lta 0.8$ data are 
from Riess \etal\ (1998) and Perlmutter \etal\ (1999).
The extreme point at $z=1.7$ indicates the upper limit for SN1997ff
from Riess \etal\ (2001).  The ``grey dust'' model represented by
the long-dashed line has a constant comoving dust density and is 
similar to the one considered by Riess \etal, except here we have 
calculated it for the open $\Omega_{\rm m}{\,=\,}0.3$,
$\Omega_{\Lambda}{\,=\,}0$ universe, rather than for an ``empty'' one.
The ``evolving dust'' models shown by the heavy ($\Omega_{\rm m}{\,=\,}0.3$)
and light ($\Omega_{\rm m}{\,=\,}1.0$) short-dashed lines have a
comoving dust density that builds up with decreasing redshift
to compensate for dilution from cosmic expansion.
The dust models are arbitrarily normalized to go through the
data points at $z{\,\approx\,}0.5$.  See text for details.
\label{fig:hubble}}
\end{figure*}

Figure~\ref{fig:hubble} places the new \sna\ distance measurements
on the differential Hubble diagram of published \sna.  
The distances are relative to a fiducial ``coasting,'' or ``empty,'' 
universe 
%(having zero matter and dark energy densities:
($\Omega_{\rm m}{=}\Omega_\Lambda{=}0$), thus removing $H_0$ from 
consideration.  The published data are
binned as in Riess \etal\ (2001) with the extreme SN1997ff point
represented by its upper limit.  The data are compared
to expectations from various models, including 
% an Einstein-deSitter model 
Einstein-deSitter ($\Omega_{\rm m} = 1$),
an open universe model ($\Omega_{\rm m} = 0.3$, $\Omega_\Lambda = 0$),
a flat $\Lambda$-dominated model 
($\Omega_{\rm m} = 0.3$, $\Omega_\Lambda = 0.7$),
and some simplistic intergalactic ``dust'' models (described below).
We use the MLCS distances for this comparison because the published
data in the figure are largely from the MLCS method or have been
homogenized to remove any mean offset from MLCS.  The data point
for SN2002dc would be offset fainter by 0.17 mag using the BAT result,
while the SN2002dd point would be virtually unchanged.

%%%%%%%%%%%%%%%%%%%%%%%%%%%%%%%%%%%%%%%%%%%%%%%%%%%%%%%%%%%%%%%%
%%%%%%%%%%%%%%%%%%%%%%%%%%%%%%%%%%%%%%%%%%%%%%%%%%%%%%%%%%%%%%%%

SN2002dc, like most individual \sna\ at similar redshifts,
is consistent with any of these models except $\Omega_{\rm m}{\,=\,}1$
without dust. SN2002dd at $z{\,=\,}0.95$
favors the $\Lambda$-dominated or dust models, but
SN1997ff at $z{\,\sim\,}1.7$
remains the sole indicator of past deceleration.
A constrained $\Omega_\Lambda+\Omega_{\rm m}\equiv1$
fit to the binned data (open symbols) gives $\Omega_\Lambda = 0.74\pm0.08$;
adding the two new MLCS distances 
and performing, purely for illustrative purposes, a hybrid binned/unbinned fit
results in $\Omega_\Lambda  =  0.73\pm0.07$.  If we replace the MLCS distances 
with our BAT results, the value becomes $\Omega_\Lambda  =  0.74\pm0.07$,
or $\Omega_\Lambda  =  0.75\pm0.07$ if the SN2002dd distance including
the rest $U$-band data is used.
Thus, our data are fully consistent with the findings of 
Riess \etal\ (1998) and Perlmutter \etal\ (1999) favoring
an accelerating universe.

However, Aguirre (1999) has argued that a uniform distribution of 
intergalactic dust can explain the faintness of
intermediate redshift \sna\ 
without invoking cosmic acceleration.  Such dust might be expected 
to have grey scattering properties because only large grains
($\gta 1 \mu$m) would likely survive the extreme conditions
required for ejection into the intergalactic medium and the
high temperatures believed to prevail there.
While the observed brightness of the $z\approx1.7$ SN1997ff argues
against the simplest, non-evolving, grey dust model and in favor
of cosmological deceleration beyond $z{\,=\,}1$ (Riess \etal\ 2001),
Goobar \etal\ (2002) show that models with an evolving
dust density can accommodate SN1997ff.

In Figure~\ref{fig:hubble}, the model labeled ``grey dust''
is very similar to the one considered by Riess \etal\ (2001), 
with the optical depth directly proportional to redshift to
approximate a fixed comoving density of dust,
except here we calculate it for the $\Omega_{\rm m} = 0.3$ open model,
rather than for $\Omega = 0$.  However, the behavior of this non-evolving
grey dust model is fairly insensitive to $\Omega_{\rm m}$.
The normalization of the extinction is adjusted to match the observed
\sna\ data near $z\approx0.5$.   The ``evolving dust'' models
shown in the figure use the same cosmology, 
as well as $\Omega_{\rm m}{\,=\,}1$, 
% and set the normalization in the same way,
but have constant {\it spatial} densities of dust,
to approximate continual injection of dust at a high enough rate
to compensate for the dilution effect of the expansion;
thus the comoving density decreases to higher redshift.  

These two types of dust model essentially reproduce the behavior
of the Goobar \etal\ (2002) models A and B, which those authors admit
are not necessarily plausible in themselves but provide
``useful limiting cases.''  The plotted ``dust'' curves can
therefore be taken to mark the approximate range of high-redshift
behavior for decelerating models which invoke grey intergalactic dust.
%to explain the faintness of intermediate-$z$ \sna.
Of course, perfectly grey dust does not exist in reality, and
efforts to detect a wavelength dependence of
the \sna\ dimming have not found any (e.g., Riess \etal\ 2000).
Moreover, a direct upper limit on the density of intergalactic
dust derived from the absence of a detectable X-ray scattering halo 
around a single distant quasar is well below that needed
to explain the dimming of \sna\ (Paerels \etal\ 2002).
%
%Therefore, cosmic acceleration seems the most natural explanation,
Therefore, cosmic acceleration seems the most natural and
compelling explanation, although only further observations
of high-$z$ \sna\ over a large wavelength range can definitively
rule out the challenge posed by these dust models.

\subsection{Systematic Effects}
% Reddening
% Lens Magnifications

Various systematic effects  on high-$z$ \sna\ distances
such as extinction, gravitational lensing, and selection bias
have been discussed in detail by Riess \etal\ (1998), Perlmutter \etal\ (1999), 
and several subsequent works.  In our case, the objects
were not found as part of any well-defined search effort,
and so a statistical correction for selection bias is not possible.
However, because they were both found well above
the detection limit (the statistical error at the discovery
epoch being only 0.01--0.02\,mag), this bias would be minimal.  We note
that previous high-$z$ \sna\ searches have not used such corrections.

Two issues are worth noting, however.  First, the fitted extinction
for SN2002dc, although fairly modest in Galactic terms, 
is large compared to most other high-$z$ \sna\ previously measured. 
It is higher than for the great majority of objects in both the
Riess \etal\ (1998) and Tonry \etal\ (2003) samples
(most of which have fitted extinctions consistent with zero),
as well as most objects from Perlmutter \etal\ (1999).
On the other hand, the uncertainty in the extinction is fairly
large, $\pm0.19$ mag from MLCS and $\pm0.30$ mag from BATM.
This is the dominant source of distance uncertainty for SN2002dc
(the extinction error for SN2002dd is about $\pm$0.16 mag
from both methods).
Still, it may be that seeing effects in ground-based data tend to 
select against \sna\ deep inside galaxies that would have
higher extinction values.
Therefore, the quality of highly extincted \sna\ distances has not been
thoroughly tested.  Further, there are no external checks of the extinction
estimate, which depends critically on the assumption that the reddening law
in the $z{=}0.47$ host is the same as that for the interstellar medium of
our Galaxy, but variations are known to occur (Cardelli \etal\ 1989;
Falco \etal\ 1999).

Another effect worth considering is weak gravitational lensing.
Stochastic lensing by large-scale structure will tend to demagnify
distant \sna\ because the clustering of galaxies
causes the typical line of sight to be underdense
(e.g., Kantowski \etal\ 1996; Wambsganss \etal\ 1997).
Both Riess \etal\ (1998) and Perlmutter \etal\ (1999)
concluded that this effect was negligible for \sna\ 
% at redshifts in the range \hbox{0.5--1.0}.
in the redshift range \hbox{0.5--1.0}.
However, certain \sna\
may be significantly magnified if their line of sight is near massive 
foreground objects.  In other words, although the magnification
probability function peaks at a value less than unity, it
is asymmetric with a tail towards higher magnification.
In the case of SN1997ff at $z{\,\approx\,}1.7$,
which is projected near several bright $z = 0.5$--0.9
galaxies, the magnification has recently been estimated at 
$-0.34\pm0.12$ mag (Ben\'\i tez \etal\ 2002), which would bring
it very close to the ``empty'' model in Figure~\ref{fig:hubble}.

We performed the same analysis for the two new \sna\ as done by Ben\'\i tez 
\etal\ (2002) for SN1997ff.  The magnification derived for SN2002dc at
$z\approx0.5$ is negligible.  However, we find a magnification
of $-0.22\pm0.10$ mag for SN2002dd, with the largest contributions coming
from galaxies 381 and 345 in the Fernandez-Soto \etal\ (1999) catalogue
(3-486.0 and \hbox{3-610.\={1}}, respectively, from Williams \etal\ 1996).
%%%
This would mean that SN2002dd is $\sim\,$10\% more distant than 
inferred above.  
Such a high magnification at $z{\,\approx\,}1$ would have a probability
of $\lta\,$0.5\%, based on the $\Lambda$-CDM galaxy-lensing
model of Holz (1998), and an even lower probability according to the nearest
equivalent model from Wambsganss \etal\ (1997).
However, it is only significant at the $\sim\,$2-$\sigma$ level,
and this analysis does not account for the demagnification
tendency of large-scale structure, which is very difficult to
estimate for individual objects. 
Thus, our magnification estimate is biased high
at some level.

To explore this issue, we repeated the 
analysis for 1000 random lines 
of sight in the HDFN and found an average magnification of $-0.05$ mag,
with 1.8\% of them yielding values as high as for SN2002dd.
If the average is taken as a zero point offset, the magnification
for SN2002dd becomes $-0.17\pm0.10$ mag.
% a more exact treatment might use the expected 
% distribution as a prior in estimating the magnification. 
%
Of course, any given estimate may have systematic errors
from the uncertain mass-to-light ratios of the
foreground galaxies and the shapes of their potentials.
Magnification corrections have not been applied to other \sna\ at $z\lta1$,
and although we also choose not to apply this correction, we note 
it is probable that SN2002dd is magnified and
the effect in Figure~\ref{fig:hubble} would be to move this
data point upward by perhaps $\sim\,$0.2\,mag.~ 
\vspace{0.2cm}

%\newpage
\section{Discussion and Summary}
\label{sec:disc}

Our analysis has added two additional high-$z$ points to the \sna\ Hubble diagram,
the first that have been found with the new ACS instrument on \hst.
Both objects were spectroscopically confirmed as \sna\ near maximum-light
by ACS grism spectra.  SN2002dd at $z{=}0.95$ has a higher redshift than 
any other spectroscopically confirmed \sna\ with a published distance.
For instance, SN1997ck at $z{=}0.97$ (Garnavich \etal\ 1998) had neither
spectroscopic confirmation nor useful color information; 
SN1997ff at $z{\approx}1.7$ also lacks spectroscopic confirmation,
has an uncertain redshift, and all the photometry was transformed to 
the $B$-band, so there was no constraint on extinction;
SN1999fv at $z{\,\approx\,}1.2$ is the most distant \sna\ with 
published confirmation (Tonry \etal\ 1999; Coil \etal\ 2000), 
but has insufficient light curve data with no color information, 
and thus only a weakly constrained distance (Tonry \etal\ 2003).
However, with the already heavy use of ACS by the \sna\ community,
we expect that this circumstance will not last for very long.

Taken at face value, our distance for SN2002dd is most consistent
with the $\Lambda$-dominated cosmology favored by the analyses of
the larger samples of ground-based \sna\ in the
$z = 0.3$--0.9 range by Riess \etal\ (1998) and Perlmutter \etal\ (1999).
Within the context of such flat models dominated by a cosmological
constant, the transition from acceleration to deceleration occurs at 
the ``coasting point'' redshift 
$z_c = [2\Omega_\Lambda/\Omega_{\rm m}]^{1/3} - 1$ (Turner \& Riess 2002).
A reliable measurement of this redshift would therefore provide a 
strong constraint on $\Omega_\Lambda$.
Although the transition occurs at $z{\,\approx\,}0.7$ for 
models with $\Lambda\approx0.7$,
the redshift of SN2002dd is still too low to yield any clear evidence
for deceleration; in particular, it remains consistent with the
simplest grey dust model.  Moreover, we estimated that SN2002dd may
be magnified by $\sim\,$0.2 mag due to the lensing effect of 
foreground galaxies, and this would push it further
into agreement with the range of low-$\Omega_{\rm m}$ dust models.
A larger sample of similar quality data on \sna\ reaching to $z\gta1.5$ 
is needed to explore fully the systematic effects, definitively test the
alternatives to acceleration, and constrain the expansion history of the universe.

In our view, the most encouraging aspect of the present work is the
relative ease with which the observations were accomplished.  Two
supernovae were found serendipitously in a single ACS frame without even
the usual search technique of image subtraction (the more distant one was
found because of its grism spectrum; later searching by image subtraction
turned up no additional bright candidates).
Both were confirmed from their spectra
as Type~Ia near maximum-light and were followed with a relatively
small expense of time, primarily with ACS.
We have also shown, using an empirical relationship,
how ACS grism data can be used to derive accurate colors and
improve light-curve coverage.  Moreover, now that NICMOS is
again functional, it is possible to sample the rest-frame visual light
of $z{\,>\,}1$ \sna, as we have done for SN2002dd.
This all bodes well for systematic searches for high-$z$ \sna\ 
using ACS; indeed such a program has already shown extraordinary
promise (Giavalisco \etal\ 2002; Riess \etal\ 2002), and we have
every expectation that ACS will continue to deliver on this promise.

\acknowledgments 

ACS was developed under NASA contract NAS 5-32864, and this research 
has been supported by NASA grant NAG5-7697 and 
by an equipment grant from  Sun Microsystems, Inc.  
The {Space Telescope Science
Institute} is operated by AURA Inc., under NASA contract NAS5-26555.
We are grateful to K.~Anderson, J.~McCann, S.~Busching, A.~Framarini, S.~Barkhouser,
and T.~Allen for their invaluable contributions to the ACS project at JHU,
and to Mark Dickinson, Marcia Rieke, and Megan Sosey for sharing their
early results on the NICMOS calibration.
We thank Brian Schmidt, the referee, for his thorough reading
of the manuscript and his many suggestions that helped to
improve the final version.

\clearpage

\begin{deluxetable}{clcccc}
% \scriptsize
%\small
\tablewidth{0pt}   % 0pt sets table to its natural width
\tablecaption{Photometry of HDFN Type~Ia Supernovae}
\tablehead{
\colhead{Epoch} &
\colhead{JD\tablenotemark{a}} &
\colhead{F775W\tablenotemark{b}} &
\colhead{F850LP\tablenotemark{b}} &
\colhead{F110W\tablenotemark{b}} &
\colhead{$k_{z}$\tablenotemark{c}} 
}
\startdata
\multicolumn{6}{c}{SN2002dc} \\ \tableline
1 & 405.4 & $22.39 \pm 0.02$ & $22.13 \pm 0.06$\tablenotemark{d}
                                                & \dots & $-$0.86 \\
2 & 415.2 &  \dots           & $22.33 \pm 0.02$ & \dots & $-$0.85 \\
3 & 438.1 & $23.55 \pm 0.03$ & $23.18 \pm 0.04$ & \dots & $-$0.92 \\
4 & 449.4 & $24.05 \pm 0.06$ & $23.45 \pm 0.07$ & \dots & $-$0.89 \\
5 & 455.2 &   \dots          & $23.82 \pm 0.06$ & \dots & $-$0.86 \\
\tableline
\multicolumn{6}{c}{SN2002dd} \\ \tableline
1 & 405.4 & $23.64 \pm 0.03$ & $23.19 \pm 0.06$\tablenotemark{d} &
                                                  \dots          & $-$1.30 \\
2 & 415.2 &   \dots          & $23.41 \pm 0.02$ & \dots          & $-$1.34 \\
3 & 438.1 & $24.98 \pm 0.06$ & $24.32 \pm 0.08$ & $23.89\pm0.10$ & $-$1.45 \\
4 & 449.4 & $25.50 \pm 0.10$ & $24.84 \pm 0.12$ & \dots          & $-$1.55 \\
5 & 455.2 &   \dots          & $24.96 \pm 0.09$ & \dots          & $-$1.57 \\
\enddata
\tablenotetext{a}{Julian Date$\,-\,$2,452,000}
\tablenotetext{b}{Vega magnitude.}
\tablenotetext{c}{$K$-correction used in transforming the observed F850LP
($z$-band) magnitude to rest-frame $R$ (SN2002dc) or $B$ (SN2002dd).}
\tablenotetext{d}{Reconstructed from the F775W magnitude
and G800L spectrum at this epoch.\ See text.}
\label{tab:mags}
\end{deluxetable}

\begin{deluxetable}{lccccccc}
%\scriptsize
%\small
\tablewidth{0pt}   % 0pt sets table to its natural width
\tablecaption{\sna\ Distances}
\tablehead{\colhead{} & \colhead{} & \colhead{} & 
%\colhead{MLCS} & \colhead{MLCS} &
\colhead{MLCS} &  &
%\multicolumn{2}{c}{MLCS} &
\colhead{} &
%\colhead{BAT} & \colhead{BAT}
%\multicolumn{2}{c}{BAT} 
\colhead{BAT} & 
\\
\colhead{SN} &
\colhead{$z$} &
\colhead{$\Delta$} &
\colhead{$(m{-}M)_0$\tablenotemark{a}} &
\colhead{$A_V$} &
\colhead{Templ.} &
\colhead{$(m{-}M)_0$\tablenotemark{a}} &
\colhead{$A_V$} 
}
\startdata
%2002dc\dotfill & 0.475 & $+$0.10 & $42.20 \pm 0.23$ & 0.18$\pm$0.19 & 1998aq & $42.37 \pm 0.35$ & 0.29$\pm$0.30 \\
2002dc\dotfill & 0.475 & $+$0.10 & $42.20 \pm 0.23$ & 0.18 & 1998aq & $42.37 \pm 0.35$ & 0.29 \\
%2002dd\dotfill & 0.950 & $-$0.40 & $44.15 \pm 0.20$ & 0.10$\pm$0.15 & 1995al & $44.27 \pm 0.27$ & 0.10$\pm$0.18 \\
2002dd\dotfill & 0.950 & $-$0.40 & $44.15 \pm 0.20$ & 0.10 & 1995al & $44.27 \pm 0.27$ & 0.10 \\
%2002dd, no $U$\tablenotemark{b}  & 0.950 & $-$0.42 & $44.17 \pm 0.20$ & 0.11$\pm$0.15 & 1990N  & $44.17 \pm 0.27$ & 0.09$\pm$0.17 \\
2002dd, no $U$\tablenotemark{b}\dots & 0.950 & $-$0.42 & $44.17 \pm 0.20$ & 0.11 & 1990N  & $44.17 \pm 0.27$ & 0.09 \\
\enddata
\tablenotetext{a}{Absolute distance moduli for a zero point with $H_0{\,=\,}65.0$ \kmsMpc.}
\tablenotetext{b}{Results omitting the F775W data (rest-frame $U$ band).}
\label{tab:results}
\end{deluxetable}

\end{document}